\documentstyle[preprint,prd,aps,epsf,epsfig]{revtex}
\begin{document}
\tightenlines
\draft
%
\title{Species Abundances Distribution in Neutral Community Models}
%
\author{Zolt\'an N\'eda$^{*}$ and M\'aria Ravasz}
\address{ Babe\c{s}-Bolyai University, Department of Theoretical Physics \\ 
          str. Kog\u{a}lniceanu 1, RO-400084 Cluj-Napoca, Romania \\ 
          $^*$  E-mail: zneda@phys.ubbcluj.ro}

\maketitle

\begin{abstract}
An analytical approximation is derived for the Zero Sum Multinomial distribution
which gives the Species Abundance Distribution in Neutral Community Models. 
The obtained distribution function describes well computer simulation results on the
model, and leads to an interesting relation between the total number of individuals,
total number of species and the size of the most abundant species of the considered
metacommunity. Computer simulations on neutral community models, proves also the validity
of this scaling relation. 
\end{abstract}

\vspace{2pc}

\section{Introduction}

Communities of ecologically similar species that compete with each other solely for resources
are often described by neutral community models (NCM) \cite{Hubbel,Bell,Chave,Norris}. These models proved to be successful and useful in describing many of the basic patterns of biodiversity such as the distribution of abundance, distribution of range, the range-abundance relation and the species-area relation \cite{Bell,Chave}. The neutral theory is considered by many ecologists as a radical shift from
established niche theories and generated considerable controversy \cite{Levine,Enquist,Abrams,Clark,Landau}. The relevance of NCM for describing the dynamics and statistics of real communities is still much debated and criticized \cite{Nee}.

Nowadays NCM are studied mostly by Monte Carlo type computer simulations \cite{Chave,Bell,McGill}, and
apparently there are no analytical results. One of the key issues that macro-ecologist are often investigating is the species abundances distribution (SAD), introduced for characterizing the frequency of species with a given abundance \cite{Preston,May,Pielou}. In case of NCM, SAD is generated numerically and it is called the Zero Sum Multinomial (ZSM) distribution \cite{McGill,Condit}.
The aim of the present paper is to give an analytical mean-field type approximation for ZSM. By using the
invariance of the system against the intrinsic fluctuations characteristic for NCM, we derive an 
analytic solution that describes the results of computer simulations. The derived
analytical form of SAD leads also an interesting relation between the total number of individuals, total
number of species and the size of the most abundant species of the considered meta-community. This
novel scaling relation is confirmed by computer simulations on neutral models. 
 
\section{Neutral Community Models}

NCM are usually defined on lattice sites, on which a given number of $S_{max}$ species can coexist
\cite{Hubbel,Bell,McGill} and compete for resources. Each lattice site can be occupied by many individuals belonging to different species, however the total number of individuals for each lattice site is limited to a fixed $N_{max}$ value. This limiting value models the finite amount of available resources in a given territory. As time passes individuals in the system can give birth to individuals belonging to the same species, can die or can migrate to a nearby site. The neutrality of the model implies that all individuals (regardless of the species they belong) are considered to be equally fit for the given ecosystem, and have thus the same $b$ multiplication, $d$ death and $q$ diffusion rate. The system is considered also in contact with a reservoir, from where with a small $w<<1$ probability per unit time an individual from a randomly chosen species can be assigned to a randomly chosen lattice site. This effect models the random fluctuations that can happen in the abundances of species. The dynamics of the considered community is than as follows:
\begin{itemize}
\item A given number of individuals from randomly chosen species are assigned on randomly chosen lattice sites.
\item With the initially fixed probabilities we allow each individual to give birth to another individual of the same species, to die or to migrate on a nearby site.
\item We constantly verify the saturation condition on each site. Once the number of individuals on a site exceeds the $N_{max}$ value, a randomly chosen individual is removed from that site. 
\item We apply the random fluctuations resulting from the reservoir.
\end{itemize}

After on each lattice site saturation is achieved a dynamical equilibrium sets in, and one can study
the statistical properties of several relevant quantities. Computer simulations usually focus on
generating SAD and on studying several scaling relations like species-area and range-abundance relations.   

\section{Analytical approximation for SAD}

Let us consider a fixed area in a NCM (a delimited region in the lattice) on which we study SAD. 
In the selected area we denote by $S(x,t)$ the number of species with size $x$ at the time-moment $t$. 
($x$ is a discrete variable $x=1,2, ....k...$). $S(x,t)$ divided by the total number of species yields the mathematically rigorously defined SAD (Species Abundances Distribution). We mention here that in most of the papers dealing with SAD, instead of this rigorously defined distribution function a histogram on
intervals increasing as a power of 2 is constructed \cite{Preston,May,Pielou}. On a log-normal scale this 
histogram usually has a Gaussian shape, and thus SAD is called a log-normal distribution.  
Without arguing on the relevance of this histogram (a nice treatment on this subject is 
given by May \cite{May}) for the sake of mathematical simplicity we will not use this representation, and calculate instead the mathematically rigorous distribution function. It is of course anytime
possible to re-plot the obtained distribution function in the form that is usually used
by ecologists, using instead of the $x$ variable the $z=log_2(x)$ variable.

In the framework of the considered model the time evolution of $S(x,t)$ for an infinitesimally short $dt$ 
time can be approximated by the following master-equation:

\begin{eqnarray}
S(x,t+dt)=S(x,t)+[W_+(x-1) S(x-1,t)+W_-(x+1)S(x+1,t)-\nonumber \\
-W_+(x) S(x,t)-W_-(x) S(x,t)]dt
\label{master}
\end{eqnarray}

In the equation from above $W_+(x)$ denotes the probability that one species with size $x$ increases its size to $x+1$ in unit time and $W_-(x)$ denotes the probability that one species with size $x$ decreases it's size to $x-1$ in unit time. We neglected here the possibility that in the small $dt$ time-interval one species increases or decreases it's size by more than one individual. The value of $dt$ can be 
always taken as small, as needed so that this starting assumption should hold. It worth also mentioning
that this master equation is not applicable in the neighborhood of the limiting values of $x$ since here either $x-1$ or $x+1$ is not existing. We expect thus that the shape of SAD determined from (\ref{master}) can have problems for very low and very high values of $x$.

We assume now that SAD reaches a steady-state in time. All computer simulations on
neutral models shows that this is true. This means that $S(x,t)$ should be time invariant in respect of the fluctuation governed by equation (\ref{master}).
Under this stationarity assumption we get the equation:

\begin{equation}
W_+(x-1) S(x-1,t)+W_-(x+1)S(x+1,t)-W_+(x) S(x,t) - W_-(x) S(x,t)=0
\label{form}
\end{equation}

We have to approximate now the $W_+(x)$ and $W_-(x)$ probabilities. We will work with
the assumptions of the NCM, and consider all species having the 
same birth, death and migration rate. Let us denote by
$P_+$ the probability that one individual multiplies itself in unit time (we assume $P_+$ is the
same for all individuals and species). Let us denote by $P_-$ the probability that one
individual disappears from the considered territory in unit time (again the same for all individuals and species). Further we assume that:
\begin{eqnarray}
P_+<<1 \nonumber\\
P_-<<1
\label{assum}
\end{eqnarray}

By simple probability theory we get:
\begin{equation}
W_+(x)= x P_+ [1-(P_++P_-)]^{x-1}
\label{w+}
\end{equation}

The above equation tells us, that the increase by unity of the size of one species can be
realized if any of the $x$ individual from a selected species 
multiplies itself, while the other individuals remain unchanged. (Of course there
are many other possibilities involving the birth and death of more than one individual. 
However, since we considered the (\ref{assum}) assumption
all other possibilities will be with orders of magnitude smaller). 
It is also worth mentioning that for the selected local community the effect of migration and the 
stochastic contribution from the reservoir can be
taken into account through the birth and death processes, changing slightly the values of this
probabilities. Migration inside the considered area is equivalent with a birth process, while migration
outside from the considered territory is equivalent with death of individuals. 

Using the assumptions (\ref{assum}) we can make now the following approximations:
\begin{eqnarray}
W_+(x)= xP_+[1-(P_++P_-)]^{x-1} = x P_+ [1-(P_++P_-)]^{[1/(P_++P_-)] \cdot (P_++P_-)(x-1)} \approx 
\nonumber \\
\approx xP_+e^{-(P_++P_-)(x-1)}
\label{am}
\end{eqnarray}

In the same manner, one can write:
\begin{equation}
W_-(x)= x P_- [1-(P_++P_-)]^{x-1} \approx x P_-e^{-(P_++P_-)(x-1)}
\label{ap}
\end{equation} 

Instead of $P_+$ and $P_-$ we introduce now two new notations:
\begin{eqnarray}
s=P_++P_- \\
q=P_+-P_- 
\end{eqnarray}

from where:
\begin{eqnarray}
P_+=\frac{s+q}{2} \\
P_-=\frac{s-q}{2}
\end{eqnarray}

From the assumptions (\ref{assum}) it is clear that it also holds:
\begin{eqnarray}
s<<1 \\
q<<1
\end{eqnarray}

Let us assume now that
\begin{equation}
|P_+|=|P_-| \longrightarrow q=0,
\end{equation}
which would mean that the probability of multiplication and death is the same, so
there is a constant number of individuals in the considered local community. In other words this means 
that the territory is saturated, and although the size of different species fluctuates, the total
number $N_t$ of population is constant.

The probability density for the species abundances distribution (SAD) is given than as:
\begin{equation}
\rho(x,t)=\frac{S(x,t)}{S_t}.
\end{equation}

Instead of $x$ let us introduce now a new variable $y=x/N_t<<1$ ($N_t>>1$ is 
the total number of individuals in the system)

For $\rho(y,t)$ we have the (\ref{form}) equation:

\begin{equation}
W_+(yN_t-1) \rho(y-\frac{1}{N_t},t)+W_-(yN_t+1) \rho(y+\frac{1}{N_t},t)-W_+(yN_t) \rho(y,t) - W_-(yN_t) \rho(y,t)=0
\end{equation} 

Since $\rho(y,t)$ is a limiting distribution (not depending on $t$ anymore) we will simply
denote is as $\rho(y)$. 

\begin{equation}
W_+(yN_t-1) \rho(y-\frac{1}{N_t})+W_-(yN_t+1) \rho(y+\frac{1}{N_t})-W_+(yN_t) \rho(y) - W_-(yN_t)  \rho(y)=0
\label{start}
\end{equation} 

We can use now Taylor series expansion to get $\rho(y-\frac{1}{N_t})$ and $\rho(y+\frac{1}{N_t})$:

\begin{eqnarray}
\rho(y-\frac{1}{N_t})=\rho(y)-\frac{1}{N_t}\rho'(y)+\frac{1}{2{N_t}^2} \rho''(y) \\
\rho(y+\frac{1}{N_t})=\rho(y)+\frac{1}{N_t}\rho'(y)+\frac{1}{2{N_t}^2} \rho''(y) 
\end{eqnarray}

We denoted here by $\rho'(y)$ and $\rho''(y)$ the first and second order derivatives
of the $\rho(y)$ function, respectively. Taking account of $q=0$, the values of $W_{\pm}(y)$ are given by (\ref{am}, \ref{ap}) as follows:

\begin{eqnarray}
W_+(yN_t-1)=\frac{(yN_t-1) s}{2} exp[-(yN_t-2)s]   \\
W_-(yN_t+1)= \frac{(yN_t+1) s}{2} exp[-yN_ts] \\
W_+(yN_t)= \frac{yN_ts}{2} exp[-(yN_t-1)s]\\
W_-(yN_t)= \frac{yN_ts}{2} exp[-(yN_t-1)s]
\end{eqnarray}

Plugging all these in equation (\ref{start}):

\begin{eqnarray}
\frac{(yN_t-1) s}{2} exp[-(yN_t-2)s] [\rho(y)-\frac{1}{N_t}\rho'(y)+\frac{1}{2{N_t}^2} \rho''(y) ]+
\nonumber \\ \frac{(yN_t+1) s}{2} exp[-yN_ts] 
[\rho(y)+\frac{1}{N_t}\rho'(y)+\frac{1}{2N_t^2} \rho''(y) ] =
\frac{yN_ts}{2} exp[-(yN_t-1)s] \rho(y)
\end{eqnarray}
Simplifying both sides with $s \cdot exp[-yN_ts]$, some immediate algebra yields the following second order differential equation for $\rho(y)$:

\begin{eqnarray}
\rho(y)[\frac{yN_t}{2}(e^{2s}-2e^s+1)-\frac{1}{2}(e^{2s}-1)] + \rho'(y) \frac{1}{N_t} [ \frac{yN_t}{2}(1-e^{2s}) + \frac{1}{2}(1-e^{2s})] + \nonumber \\ 
+\rho''(y) \frac{1}{2N_t^2} [(\frac{yN_t}{2} (e^{2s}+1)+\frac{1}{2}(1-e^{2s})]=0
\end{eqnarray}

Since $s<<1$ the following approximations are justified
\begin{eqnarray}
e^{2s} \approx 1+2s \\
e^{s} \approx 1+s ,
\end{eqnarray}

and the differential equation becomes:
\begin{equation}
-\rho(y) s + \rho'(y) \frac{1}{N_t} [s+1-yN_ts] + \rho''(y) \frac{1}{2N_t^2} [yN_t+yN_ts-s] = 0  
\end{equation}

For solving this differential equation, in the {\bf first approximation} we neglect 
all term that are proportional with the $1/N_t \rightarrow 0$ quantity. This yields
a first order differential equation:

\begin{equation}
s \rho(y) = -y s \rho'(y)
\end{equation}

This equation has the immediate solution

\begin{equation}
\rho_I(y)=C_1/y,
\label{app1}
\end{equation}

with $C_1$ an integration constant.
 
The histogram $\sigma (z)$ that is usually used for SAD can be immediately determined from (\ref{app1}),
writing the $\rho_I(y)$ distribution as a function of the $z=log_2(x)=log_2(yN_t)$ variable. It is immediate to realize that this would yield a constant distribution ($\sigma_{I}(z)=C$). 

A {\bf better approximation} can be achieved by keeping the terms proportional with $1/N_t$ and
neglecting the second orderly small $1/N_t^2$ and $s/N_t$ terms. This yields the 

\begin{equation}
-s \rho(y) + \rho'(y) \frac{1}{N_t} [1-yN_ts] + \rho''(y)\frac{1}{2N_t} y =0 
\end{equation}

differential equation. Going back now to the $x=yN_t$ variable

\begin{equation}
-s \rho(x) + \rho'(x) [1-xs] + \rho ''(x) \frac{x}{2} =0,
\end{equation}
we get the general solution

\begin{equation}
\rho_{II}(x)=\frac{C_1}{x}+\frac{e^{2sx} C_2}{x},
\end{equation}

where $C_1$ and $C_2$ are two integration constants.

By visually comparing with the experimental and simulated SAD curves we can conclude that 
we need $C_1>0$ and $C_2<0$ to get the right shape. The general solution for SAD, should write thus

\begin{equation}
\rho_{II}(x)= \frac{K_1}{x}(K_2-e^{2sx}),
\end{equation}
with $K_1$ and $K_2$ two real, positive constants. 

It is immediate to observe that the obtained distribution for SAD, has a cutoff, i.e. there
is a maximum value of $x$ until $\rho(x)$ is acceptable (remains positive). This results, is not surprising, since due to the finite number of individuals in the system and the finite value of 
the number of species one would naturally expect a cutoff in the distribution. 

There are three fitting parameters in the mathematical expression of $\rho_{II}(x)$ ($K_1$, $K_2$ and $s$). Since $\rho_{II}(x)$ has to be normalized, we can 
determine $K_1$ as a function of $K_2$ and $s$. The normalization of this distribution function
is not easy and cannot be done analytically, since there is no primitive function for
$exp(\alpha x)/x$.  
 
However, if we can use the $sx<<1$ assumption and consider a Taylor expansion 
in the exponential we obtain the more simple
\begin{equation}
\rho_{II}(x) \approx F_n \frac{F_1-x}{x},
\label{form2}
\end{equation}
($F_n$ and $F_1$ are again two positive real constants) distribution, which has a cutoff 
for $x=F_1$.  
This distribution function is exactly the same as the one proposed by Dewdney using 
totally different arguments \cite{Dewdney},
and named {\em logistic-J distribution}. As argued in \cite{Dewdney} it describes well
the SAD for many real communities.

The normalization condition for this distribution function is:
\begin{equation}
\int_{1}^{F_1} F_n \frac{F_1-x}{x} dx =1,
\label{normal}
\end{equation}
and an immediate calculus gives:
\begin{equation}
F_n=\frac{1}{F_1 ln(F_1)-F_1+1},
\end{equation}

The approximated normalized distribution function for SAD is then:

\begin{equation}
\rho_{II}(x) \approx \frac{1}{F_1 ln(F_1)-F_1+1} \frac{F_1-x}{x}
\label{result}
\end{equation} 

We can consider thus the above simple one-parameter fit to approximate the results for SAD on NCM. 

The shape of $\sigma(z)$ can be again quickly obtained
from $\rho_{II}(x)$, by changing the variable in this distribution function to $z=log_2(x)$.
A simple calculation yields the form
\begin{equation}
\sigma_{II}(z)=C*(F_1-2^z),
\label{app2}
\end{equation}
where C is another normalization constant. It is important to realize, that $\sigma(z)$ 
given by the above approximation does not show the generally observed bell shaped curve,
and for small values of $z$ it is a constant. We must remember however that the shape of SAD given by our approximation can not be trusted for small $z$ values, since in this limit the starting master equation (\ref{master}) is not valid.   

\section{SAD from computer simulations}

In order to check the validity of our analytical approximation for SAD we performed computer simulations
on the model presented in Section 2. We considered a lattice of size $20 \times 20$, $S_t=400$ species, and $N_{max}=1000$ for each lattice site. We studied a local community on a square of 
$9\times 9$ lattice sites, and we fixed several values for the dynamical parameters $d/b$ and $q/b$. 
We used periodic boundary conditions, and the efficient kinetic or resident time Monte Carlo 
algorithm was implemented. The simulations were made on a $Pentium^{(TM)} 4$ cluster. As a general results, we obtained that the analytical form given by (\ref{result}) describes well the simulation data for SAD. On Figure 1 we present a characteristic fit for the simulation data. The parameters 
used in the simulation were $d/b=0.3$ and $q/b=0.2$. The obtained best fit parameter for 
equation (\ref{result}) was $F_1=14500$. The rigorously 
defined $\rho(x)$ distribution function suggest that in NCM SAD has a scale-invariant nature. The finite size of the system introduces a natural cutoff in this
scale-invariant behavior.   
  
Computer simulations on NCM proves thus the applicability of our analytical approximations for the
form of the ZSM distribution.

\section{Scaling laws resulting from SAD}

Starting from the analytical approximation (\ref{result}) for the form of SAD, we can derive
an interesting relation between the size of the most abundant species ($N_s$), the total number
of individuals ($N_t$) and the number of detected species ($S_t$) in the considered meta-community. 
The distribution function (\ref{result}) has a cutoff at $x=F_1$, from where it results that
$F_1 \approx N_s$. It is also immediate to realize that from the definition of $\rho(x)$ it results
\begin{eqnarray}
N_t=\int_{1}^{N_s} C x \frac{\rho(x)}{F_n} dx = C \int_{1}^{N_s} x \frac{N_s-x}{x} dx \\
S_t=\int_{1}^{N_s} C \frac{\rho(x)}{F_n} dx = C \int_{1}^{N_s} \frac{N_s-x}{x} dx, 
\end{eqnarray} 
where $C$ is a normalization constant, which normalizes $\rho(x)$ to the total number of 
species in the local community.  The above two integrals are easily calculated and
leads to the following two coupled differential equations:
\begin{eqnarray}
N_t=C N_s (N_s-1) - \frac{C}{2} (N_s^2-1) \\
S_t=C N_s ln(N_s) - C (N_s-1)
\end{eqnarray}
Working on relatively large habitats, one can use the $N_s>>1$ assumption, and the coupled equation system from above can be simplified:
\begin{eqnarray}
N_t \approx \frac{C}{2} N_s^2 \\
S_t \approx C N_s [ln(N_s)-1] 
\end{eqnarray}
Eliminating from this system the normalization constant $C$ we obtain the important relation:
\begin{equation}
\frac{S_t N_s}{N_t [ln(N_s)-1]}=2
\label{magic}
\end{equation}

Computer simulation results on NCM supports again the validity of the magic formula from above.
(The simulations were made on a $20 \times 20$ lattice, and we choose $S_t=400$, $N_{max}=1000$, $d/b=0.3$ and $q/b=0.2$).
On Figure 2 we plotted the simulation results for different local community sizes, and 
the plot shows that equation (\ref{magic}) works well, however the constant on the right
side of the equation seems to be slightly different from 2. 
We think that this slight difference is the result of our crude approximation: $F_1 \approx N_s$, and in reality we should have $F_1$ slightly bigger than $N_s$. The simulation data from Figure 2
was obtained after averaging on several local communities of size $A$.

Increasing the size $A$ of the considered habitat one would naturally expect $N_t \sim A$. Using
equation (\ref{magic}) one would immediately get thus the interesting scaling-law:

\begin{equation}
\frac{S_t N_s}{ln(N_s)-1} \sim A
\label{scaling}
\end{equation}
  
The (\ref{scaling}) scaling relation can be also immediately verified in computer simulations
on NCM. Results for a $20 \times 20$ lattice, $S_t=400$, $N_{max}=1000$, $d/b=0.3$ and $q/b=0.2$ are shown on Figure 3. On the figure with a dashed line we indicated the power-law with exponent $1$. 
As seen from the figure, the simulation data supports the scaling-law given by our analytical
approach. 
\section{Conclusions}

We have given here an mean-field type analytical approximation for the species abundances distribution function for neutral community models. By using the invariance of this distribution regarding the internal
fluctuations characteristic for the model, we derived an analytical approximation for the
distribution function which describes well the simulation data obtained on NCM. The derived distribution
function has a natural cutoff, governed by the finite extent of the system, and leads to
an interesting relation between the total number of individuals, total number of species and the size of the most abundant species, found in the considered habitat. Computer simulations on neutral models
confirms the validity of this scaling relation.  
\section{Acknowledgments}

The present study was supported by the Sapientia KPI foundation for interdisciplinary research.
We  are grateful for Dr. N. Stollenwerk for helpful suggestions and discussions. We also
thank Dr. A. Balogh and Dr. V. Mark\'o for introducing us in this fascinating interdisciplinary field,
and for providing us a lot of interesting bibliography on the subject.

\begin{figure}
    \begin{center}
    \epsfig{figure=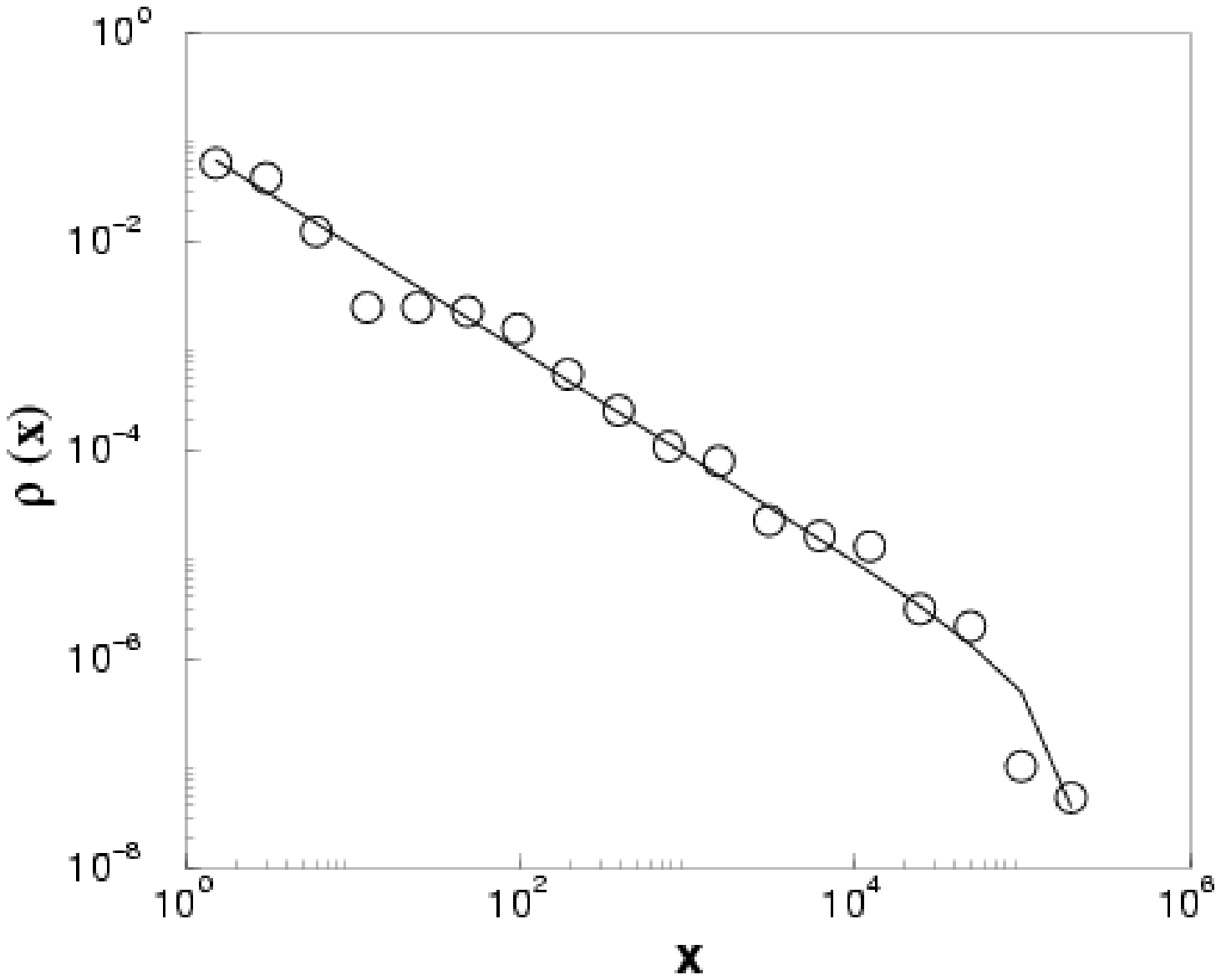,width=10.0cm,angle=0}
    \caption{Characteristic fit (continuous line) with equation (\ref{result}) for 
    the simulation results (circles) on Species Abundances Distribution in Neutral Community Models.  
    Simulations done on a 
    $20 \times 20$ lattice, $S_t=400$, $N_{max}=1000$, $d/b=0.3$ and $q/b=0.2$. The 
    best fit parameter for SAD yield $F_1=14500$.}
    \end{center}
\end{figure}

\begin{figure}
    \begin{center}
    \epsfig{figure=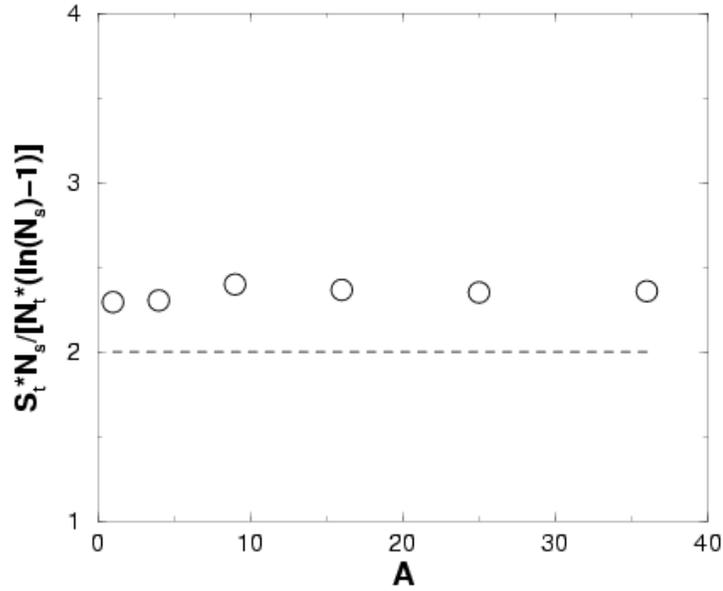,width=10.0cm,angle=0}
    \caption{Validity of the (\ref{magic}) equation. Circles are simulation data and the
dashed line indicates the expected value of $2$. Parameters of the simulation are the same
as in Figure 1. }
    \end{center}
\end{figure}

\begin{figure}
    \begin{center}
    \epsfig{figure=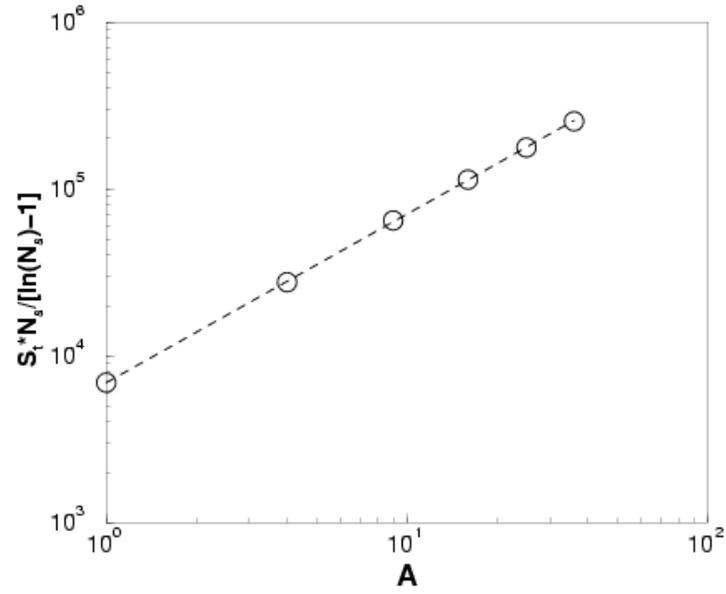,width=10.0cm,angle=0}
    \caption{Validity of the (\ref{scaling}) scaling law. Circles are simulation data and the
dashed line indicates a power-law with exponent $1$. Parameters of the simulation are the same as in
Figure 1.}
    \end{center}
\end{figure}

\end{document}